\documentclass[useAMS,usenatbib]{mn2e}
\usepackage{txfonts}
\usepackage{natbib}
\usepackage[all]{xy}

\usepackage[dvips]{graphicx}
\def\del#1{{}}

\newcommand{\ltsima}{$\; \buildrel < \over \sim \;$}
\newcommand{\lsim}{\lower.5ex\hbox{\ltsima}}
\newcommand{\gtsima}{$\; \buildrel > \over \sim \;$}
\newcommand{\gsim}{\lower.5ex\hbox{\gtsima}}
\newcommand{\bra}{\langle}
\newcommand{\ket}{\rangle}

\newcommand{\dd}{\mathrm{d}}

\title[Spherical collapse in modified gravity]
{Spherical collapse in modified gravity with the Birkhoff-theorem}
\author[Bj{\"o}rn Malte Sch{\"a}fer and Kazuya Koyama]
{Bj{\"o}rn Malte Sch\"afer\thanks{e-mail: Bjoern.Schaefer@port.ac.uk} and Kazuya Koyama\\
Institute of Cosmology and Gravitation, University of Portsmouth, Mercantile House, Hampshire Terrace, Portsmouth PO12EG, United Kingdom}

\begin{document}
\pagerange{\pageref{firstpage}--\pageref{lastpage}}
\pubyear{2003}
\maketitle
\label{firstpage}

% --- abstract --- %
\begin{abstract}
We study structure formation in a phenomenological model of modified gravity which interpolates between $\Lambda$CDM and phenomenological DGP-gravity. Generalisation of spherical collapse by using the Birkhoff-theorem along with the modified growth equation shows that the overdensity for spherical collapse $\delta_c$ in these models is significantly lowered compared to $\Lambda$CDM, leading to enhanced number densities of massive clusters and enhanced cluster merging rates. We find that $\delta_c(z)$ is well fitted by a function of the form $\delta_c(z) = a - b\exp(-cz)$. We examine the sensitivity of PLANCK's and SPT's Sunyaev-Zel'dovich survey to constrain the modified gravity parameterisation and find that these experiments can easily distinguish between models with a cosmological constant and modified gravity, if prior constraints from CMB temperature and polarisation anisotropies are included.
\end{abstract}

% --- keywords --- %

% --- section: introduction --- %
\section{Introduction}
The late-time accelerated expansion of the universe, observed in the magnitude redshift relation of type Ia supernovae, can be explained by three very different theories, either by assuming that the expansion is due to the cosmological constant $\Lambda$, or due to dark energy as an additional fluid, or due to a modification in the gravitational law. Although there has been rapid progress in developing modified theories of gravity to explain the late-time accelerated expansion of the universe, none of the models is yet convincing. The Dvali-Gabadadze-Porrati (DGP) model \citep{2000PhLB..485..208D} for example, which is based on a five-dimensional brane world set-up, offers a simple explanation of the accelerated expansion of the universe from 5d physics without dark energy, but it has been shown that the model suffers from a ghost instability \citet{2007arXiv0709.2399K}. Simple modifications of the 4d Einstein-Hilbert action such as $f(R)$ gravity violate solar system constraints unless $f(R)$ is fine-tuned \citep{2003PhLB..575....1C}. Structure formation with DGP-gravity was investigated by \citet{2004PhRvD..69d4005L, 2004PhRvD..69l4015L}, and constraints on DGP gravity have been derived (CMB: \citet{2007PhRvD..75f4003S}, supernovae: \citet{2007JCAP...05....3R}, large-scale structure: \citet{2006PhRvD..74b3004M, 2006JCAP...01..016K}; a review is provided by \citet{2007arXiv0706.1557K}).

However, we have learnt general features of modified gravity from these attempts: Firstly, any modification of general relativity (GR) on cosmological (horizon) scales results in modifications of GR on sub-horizon scales as well, which influences the growth of perturbations. Whereas in cosmologies with homogeneous dark energy the growth function is uniquely determined by the Hubble-function, modified gravity models generically alter the growth function and it becomes impossible to reconcile measurements of the expansion and growth rates resulting from a modified gravity model with a dark energy model with standard gravity. This leads to the interesting possibility of distinguishing the modified gravity models from dark energy models using probes of the cosmological large-scale structure such as the CMB or the integrated Sachs-Wolfe effect. Secondly, any successful modified gravity models should satisfy the stringent constraints coming from the local tests of gravity. This effectively means that the theory must approach GR on small scales. The DGP-model naturally recovers GR on small scales without introducing any additional phenomena due to the strong coupling. An example of this is $f(R)$-gravity, where a very contrived form of the $f(R)$-function (the so-called Chameleon mechanism) is necessary to recover GR \citep{2007arXiv0708.1190H}. 

An interesting probe of the growth of structure is provided by weak lensing, but the non-linear transition to GR is important for the correct calculation of the weak lensing signal. The common approach in the literature is to parameterise the linear growth rate and to use the non-linear mapping of the linear matter power spectrum developed in GR to calculate nonlinear growth and the nonlinear lensing signal. This is effectively equivalent to assuming that gravity is modified even on small scales which would contradict the local gravity experiments. Recently, attempts have been made to carry out $n$-body simulations in modified gravity \citep{2006PhRvD..74h4007S, 2007arXiv0709.0307L}, but the mechanism to recover GR due to non-linear dynamics was not properly implemented. An interesting phenomenological attempt was made to address this non-linear transition to GR using the halo-model \citep{2007arXiv0708.1190H}, but it is still an open question to calculate the non-linear power spectrum.

In this paper we investigate constraints on modified gravity as an alternative to dark energy \citep{2003astro.ph..1510D, 2006JCAP...03..017K} from structure formation and from the abundance of clusters of galaxies by solving the homogeneous growth and the spherical collapse equations. As the model for modified gravity we employ a phenomenological model which is able to interpolate between $\Lambda$CDM and DGP-gravity, and predict the number density of clusters of galaxies in a Press-Schechter approach. Due to the spherical symmetry, it is possible to address the non-linear dynamics of the gravity in a tractable way. Similarly to the case of dark energy \citep{1991MNRAS.251..128L, 2003MNRAS.344..761M}, cluster counts are a promising tool for investigating modified gravity as the collapse threshold for objects forming at late times is lowered, which gives rise to an enhanced spatial density of clusters. The influence of dark energy on cluster formation was extensively studied \citep[][ to name but a few]{1998ApJ...508..483W,2005A&A...443..819P, 2006A&A...454...27B, 2005JCAP...07..003M}, and we aim to forecast constraints on modified gravity from cluster count experiments.

This paper is structured as follows: The theory is developed in Sect.~\ref{sect_theory}, where our model for modification of gravity is outlined. The results are compiled in Sect.~\ref{sect_homogeneous} for homogeneous cosmology, in Sect.~\ref{sect_collapse} concerning the solution of the spherical collapse equations, and in Sect.~\ref{sect_number_counts} the impact of modified gravity on cluster number counts and merging rates is discussed. Constraints on cluster formation with modified gravity resulting from future Sunyaev-Zel'dovich surveys are given in Sect.~\ref{sect_szsurvey}. A summary in Sect.~\ref{sect_summary} concludes the paper. Parameter choices for the fiducial cosmological model are based on the spatially flat $\Lambda$CDM cosmology with adiabatic initial perturbations: Specific choices for the cosmological parameters are $\Omega_m=0.25$, $H_0=100\: h\:\mathrm{km}\:\mathrm{s}^{-1}\:\mathrm{Mpc}^{-1}$ with $h=0.72$, $\Omega_b=0.04$, $n_s=1$ and $\sigma_8=0.8$.

% --- section: structure formation in modified gravity --- %
\section{Structure formation in modified gravity}\label{sect_theory}
\subsection{Phenomenological DGP-gravity}
A lack of satisfying theoretical models for modified gravity enforces us to make phenomenological approaches. We assume that the modified Friedmann equation (normalised to $H_0$) is given by 
\begin{equation}
H^2(a) = (1-\Omega_m) H^\alpha(a) + \frac{\Omega_m}{a^3},
\label{eqn_friedmann}
\end{equation}
where $\alpha$ smoothly interpolates between $\Lambda$CDM ($\alpha=0$) and DGP-gravity ($\alpha=1$) \citep{2003astro.ph..1510D}. We restrict ourselves to linear structure formation as well as on object formation by spherical collapse, where we incorporate modified gravity in the growth equation and use the Birkhoff-theorem for describing the spherical collapse dynamics instead of solving the structure formation equations for full DGP-gravity.

\citet{2006JCAP...03..017K} show that a covariant effective theory that reproduces eqn.~(\ref{eqn_friedmann}) can be constructed for $\alpha=1/n$ where $n$ is an integer number. Although the Bianchi-identity would constrain the structure of the theory, we need to supply additional information in order to derive the equations that govern the evolution of structure. In this paper, we choose the Birkhoff's theorem as a guidance to study the evolution of structure which ascertains that the metric observed by a test particle outside a spherically symmetric matter distribution is equivalent to that of a point source of the same mass located at the centre of the sphere \citep{2004PhRvD..69l4015L}. Recently, it has been argued that the violation of Birkhoff's theorem poses severe conceptual and computational difficulties \citep{2007arXiv0709.4391D}: If the gravitational force in the interior of a spherically symmetric matter distribution is non-zero, one would need to include the details of the host structure a galaxy is residing in when determining gravitational forces inside a galaxy. We should note that the DGP-gravity does {\it not} respect Birkhoff's theorem. Nevertheless, a Birkhoff-like cancellation of the gravitational acceleration inside a spherical mass is enforced from the five-dimensional physics. Thus, the Birkhoff's theorem is surely not the unique way to avoid the above described conceptual difficulties, but it should give reasonable approximation to the behaviour of gravity in viable modified gravity models. In fact it was shown that the difference between using Birkhoff's theorem and the DGP-like gravity is small in general cosmologies described by eqn.~(\ref{eqn_friedmann}) concerning linear growth rate of structure \citep{2006JCAP...03..017K}.

\subsection{Birkhoff's theorem}
In this section, we explain how the Birkhoff's theorem can be used to describe the growth of overdensities. The modified Friedmann equation can be rewritten in the form
\begin{equation}
H^2= H_0^2 g(\xi_0), \quad \xi_0 = \frac{8 \pi G \rho }{3 H_0^2}.
\label{Friedmann2}
\end{equation}
Taking the time derivative of the Friedmann equation, we obtain
\begin{equation}
\frac{\ddot{a}}{a} = H_0^2 \left[   
g(\xi_0) -\frac{3}{2} \xi_0 g'(\xi_0)
\right],
\label{eqn_friedmann_density}
\end{equation}
where $g'(\xi)=d g(\xi)/d \xi$. Let us consider a spherically symmetric collapsing dust shell with radius $r(t)$. If  Birkhoff's theorem holds, the dynamics of $r(t)$ can be determined as 
\begin{equation}
\frac{\ddot{r}(t)}{r(t)} = H_0^2 \left[    
g(\xi) - \frac{3}{2} \xi g'(\xi)
\right],
\label{eqn_spherical}
\end{equation}
where 
\begin{equation}
\xi = \frac{8 \pi G \rho_\mathrm{shell}}{3 H_0^2}, \quad 
\rho_\mathrm{shell} = \frac{3 M}{4 \pi r(t)^3}.
\end{equation} 
The overdensity is defined by 
\begin{equation}
\delta = \frac{\rho_\mathrm{shell} -\rho}{\rho}.
\end{equation}
Initially, the shell is expanding due to the expansion of the Universe such that initially $r(t)=a(t) r_0$ and $\rho_\mathrm{shell}=\rho$. The conservation of mass in the shell means that $r(t)$ and $\delta$ are related by the condition $r^3(t) \rho_\mathrm{shell} = a^3 r_0^3 \rho$ and it is possible to derive the evolution for $\delta$ from the equation for $r(t)$. The linearised evolution equation for $\delta\ll 1$ is then given by
\begin{equation}
\ddot{\delta} + 2 H \dot{\delta} = 4 \pi G\Big[   
g'(\xi_0) + 3 \xi_0 g''(\xi_0) \Big] \rho \delta.
\label{eqn_lineardensity}
\end{equation}
Eqns.~(\ref{eqn_spherical}) and (\ref{eqn_lineardensity}) are the basic equations that describe the growth of structure in this model.

% --- section: homogeneous cosmology --- %
\section{Homogeneous cosmology}\label{sect_homogeneous}
\begin{figure}
\resizebox{\hsize}{!}{\includegraphics{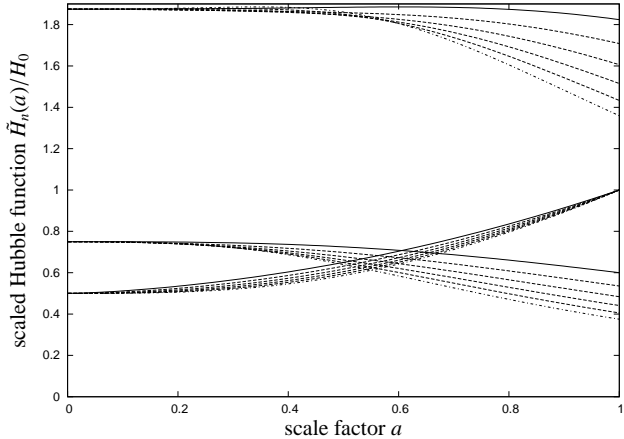}}
\caption{Hubble function $H(a)$ and its first and second derivatives, in the notation $\tilde{H}_n = a^{-\frac{2n+3}{2}}\:\dd^n H/\dd a^n$, $n=0,1,2$, for $\alpha=0$ (corresponding to $\Lambda$CDM, dash-dotted), $\alpha=1$ (DGP-gravity, solid) and intermediate values in steps of $\Delta\alpha=0.2$ (dashed).}
\label{fig_hubble}
\end{figure}

The Hubble function $H(a)$ follows from the Friedmann equation numerically by solving eqn.~(\ref{eqn_friedmann}) as an algebraic equation with the parameter $a$. All higher derivatives follow analytically by differentiating both sides of the Friedmann equation and isolating the respective derivative, e.g. for $\dd H/\dd a$ one obtains:
\begin{equation}
\frac{\dd H}{\dd a} = \frac{3\Omega_m}{a^4}\left(\alpha (1-\Omega_m) H^{\alpha-1} - 2H(a)\right)^{-1}.
\end{equation}
For visualisation of the Hubble function and its derivatives, the respective SCDM scaling is divided out:
\begin{equation}
\tilde{H}_n(a) = a^{-\frac{2n+3}{2}}\:\frac{\dd^n H}{\dd a^n}.
\end{equation}
The results for $\tilde{H}_n(a)$ are depicted in Fig.~\ref{fig_hubble}. The effect of modifying gravity according to eqn.~(\ref{eqn_friedmann}) is a more gradual increase of the Hubble function $\tilde{H}_0(a)$ compared to $\Lambda$CDM, which is reflected by smaller values of the derivative $\tilde{H}_1(a)$ and a smaller curvature $\tilde{H}_2(a)$.

\begin{figure}
\resizebox{\hsize}{!}{\includegraphics{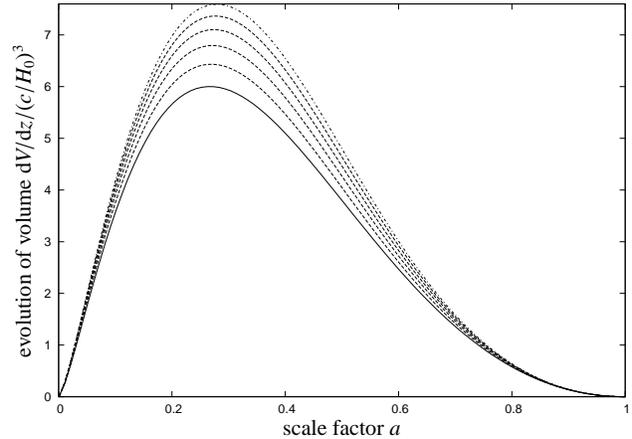}}
\caption{Evolution $\dd V/\dd z$ of comoving volume $V$ with redshift $z$, as a function of scale factor $a$, for $\alpha=0$ (corresponding to $\Lambda$CDM, dash-dotted), $\alpha=1$ (DGP-gravity, solid) and intermediate values in steps of $\Delta\alpha=0.2$ (dashed).}
\label{fig_dvdz}
\end{figure}

With the Hubble function $H(a)$, the angular diameter distance in units of the Hubble distance $c/H_0$ can be defined:
\begin{equation}
d_A(a) = a\int_a^1\dd a\:\frac{1}{a^2 H(a)},
\end{equation}
which yields the evolution of comoving volume with redshift, in units of the Hubble distance cubed $(c/H_0)^3$, for $4\pi$ solid angle:
\begin{equation}
\frac{\dd V}{\dd z}(a) = 4\pi \frac{d_A^2(a)}{a^2 H(a)}.
\end{equation}
The evolution $\dd V/\dd z$ of comoving volume as a function of scale factor $a$ is given in Fig.~\ref{fig_dvdz}. The models vary most at $a\simeq0.3$, where they exhibit a variation of almost 20\%. The $\Lambda$CDM model shows the most dramatic increase of comoving volume as a function of redshift, which is due to the rather fast transition between the matter dominated and $\Lambda$-dominated phases.

\begin{figure}
\resizebox{\hsize}{!}{\includegraphics{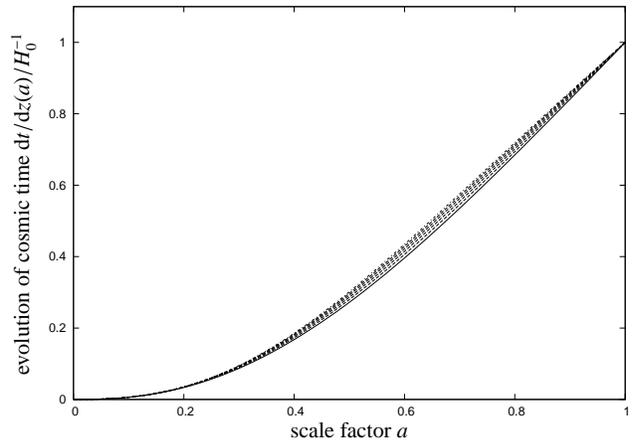}}
\caption{Evolution $\dd t/\dd z$ of cosmic time $t$ with redshift $z$, as a function of scale factor $a$, for $\alpha=0$ (corresponding to $\Lambda$CDM, dash-dotted), $\alpha=1$ (DGP-gravity, solid) and intermediate values in steps of $\Delta\alpha=0.2$ (dashed).}
\label{fig_dtdz}
\end{figure}

The evolution of cosmic time with redshift, in units of the Hubble time $1/H_0$, follows directly from the definition of the Hubble function as the logarithmic derivative of scale factor $a$ with cosmic time $t$, $H=\dd\ln a/\dd t$, and the relation between redshift $z$ and scale factor $a$, $a=1/(1+z)$:
\begin{equation}
\frac{\dd t}{\dd z}(a) = \frac{a}{H(a)}.
\end{equation}
The evolution of cosmic time $\dd t/\dd z$ as a function of scale factor $a$ is plotted in Fig.~\ref{fig_dtdz}. The influence of modified gravity seems to be minor, amounting to a few percent at $a=0.6$. Similar to the case of $\dd V/\dd z$, the modified gravity models show a slower evolution $\dd t/\dd z$ compared to $\Lambda$CDM. Appendix~\ref{sect_appendix_eos} contains a detailed comparison of modified gravity models with dark energy models with identical expansion histories, focusing on the effective dark energy equation of state $w(a)$.

\begin{figure}
\resizebox{\hsize}{!}{\includegraphics{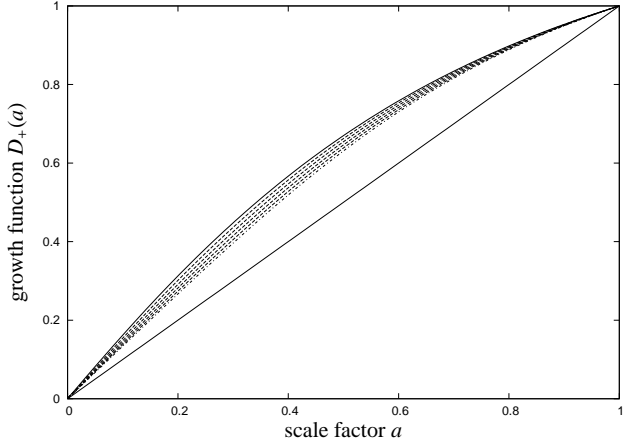}}
\caption{Growth function $D_+(a)$, for $\alpha=0$ (corresponding to $\Lambda$CDM, dash-dotted), $\alpha=1$ (DGP-modified gravity, solid) and intermediate values in steps of $\Delta\alpha=0.2$ (dashed). Additionally, the growth function $D_+(a)=a$ for the SCDM cosmology was plotted.}
\label{fig_dplus}
\end{figure}

The last point is the growth equation, which follows from the linearised structure formation equations and describes the linear and homogeneous growth of the cosmic density field with time, $\delta(\bmath{x},a) = D_+(a)\delta(\bmath{x},a=1)$. The growth equation with the scale factor $a$ as the time variable reads (c.f. eqn~\ref{eqn_lineardensity}):
\begin{equation}
\frac{\dd^2}{\dd a^2}D_+ + \frac{1}{a}\left[3+\frac{\dd\ln H(a)}{\dd\ln a}\right]\frac{\dd}{\dd a}D_+ =
\frac{3}{2a^2} G(a)\Omega_m(a) D_+(a).
\label{eqn_growth}
\end{equation}
The modification in the gravitational law is in corporated in the function $G(a)$ \citep{2004PhRvD..69d4005L}:
\begin{equation}
G(a) = \frac{2 a^4}{3\Omega_m}\left(3 H(a) \frac{\dd H}{\dd a} + a \left(\frac{\dd H}{\dd a}\right)^2 + a H(a) \frac{\dd^2 H}{\dd a^2}\right),
\end{equation}
and is reflected in the epoch-dependence of the matter density parameter $\Omega_m(a)$,
\begin{equation}
\Omega_m(a) = \frac{\Omega_m}{a^3 H^2(a)}.
\label{eqn_omegam}
\end{equation}
Fig.~\ref{fig_dplus} shows the growth function $D_+(a)$ resulting from numerical integration with initial conditions $D_+=0$ and $\dd D_+/\dd a=1$ at $a=0$, with $\alpha$ as a free parameter. The normalisation is chosen such that $D_+=1$ today. Common to modified gravity models is a faster growth of structure at early times, followed by a slower structure growth from the point on where the modification of gravity becomes important. The largest spread between the models with different $\alpha$ can be observed in the vicinity of $a=0.4$, where the growth functions differ by almost 10\%. Appendix~\ref{sect_appendix_growth} provides a detailed discussion about the magnitude of the terms governing the evolution of the growth equation eqn.~(\ref{eqn_growth}).

% --- section: spherical collapse --- %
\section{spherical collapse}\label{sect_collapse}
\begin{figure}
\resizebox{\hsize}{!}{\includegraphics{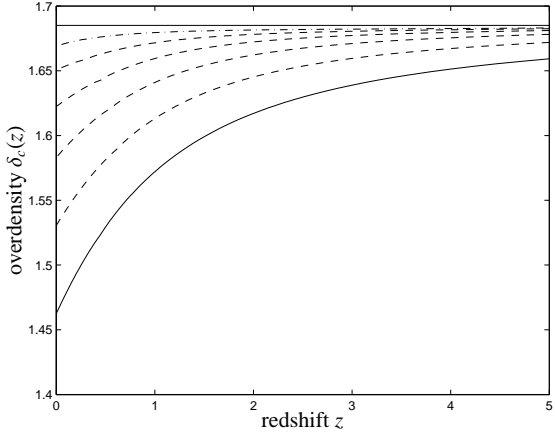}}
\caption{Overdensity for spherical collapse $\delta_c(z)$, for $\alpha=0$ (corresponding to $\Lambda$CDM, dash-dotted), $\alpha=1$ (DGP-gravity, solid) and intermediate values in steps of $\Delta\alpha=0.2$ (dashed). Additionally, the constant value of $\delta_c=1.686$ is plotted for spherical collapse in the SCDM cosmology.}
\label{fig_deltac}
\end{figure}

Using the notation introduced by \citet{1998ApJ...508..483W} and \citet{2006A&A...454...27B}, all relevant quantities are normalised by their values at turn-around, i.e. the scale factor $a$ and the radius $r$ of the density perturbation become
\begin{equation}
x\equiv\frac{a}{a_a},\quad y\equiv\frac{r}{r_a}.
\end{equation}
Generalising the Friedmann equations for incorporating the modification of gravity on large scales outlined above yields (compare eqns.~\ref{eqn_spherical} and~\ref{eqn_friedmann_density})
\begin{eqnarray}
\dot{x} & = & \sqrt{x^2 g\left(\frac{\omega}{x^3}\right)},\quad\mathrm{and}\label{eqn_xcollapse}\\
\ddot{y} & = & y\left[g\left(\frac{\omega\zeta}{y^3}\right) -\frac{3}{2}\frac{\omega\zeta}{y^3}g^\prime\left(\frac{\omega\zeta}{y^3}\right)\right],\label{eqn_ydotdot}
\end{eqnarray}
with $\omega=\Omega_m(a_a)$ and $\lambda=1-\omega$, assuring flatness. In this equation, a perturbation with overdensity $\zeta$ (measured at turn-around) will collapse at a specified time. $\zeta$ can be computed by solving the differential equation defining $y(\tau)$ subject to the initial condition $y(\tau=0)=0$ and the boundary condition $y=1$ at turn-around, which is characterised by $\dot{y}=0$. At early times, $g(x)\simeq x$ and the collapse equations reduce to the SCDM case,
\begin{equation}
\dot{x} = \sqrt{\frac{\omega\zeta}{x}},\quad\mathrm{and}\quad
\ddot{y} = -\frac{\omega\zeta}{2y^2}.
\end{equation}
with the dot denoting derivatives with respect to the dimensionless time parameter $\tau$,
\begin{equation}
\tau\equiv H(a_a)t.
\end{equation}
The solution for $x(\tau)$ can only be found numerically in the general case\footnote{An exact solution for the $\Lambda$CDM case ($\alpha=0$) can be derived in terms of the hypergeometric function, $\tau(x)=\frac{2}{3\sqrt{\omega}}x^{\frac{3}{2}} \vphantom{F}_2F_1(\frac{1}{2},\frac{1}{2},\frac{3}{2},-\frac{\lambda}{\omega}x^3)$.}, but simplifies to the SCDM result at early times $x\ll 1$ or for $\lambda=0$:
\begin{equation}
\tau = \frac{2}{3}\frac{x^\frac{3}{2}}{\sqrt{\omega}}.
\label{eqn_xsolution}
\end{equation}
The integration of $\ddot{y}$ involves multiplying both sides with $\dot{y}y$ and noticing that
\begin{equation}
\frac{1}{2}\frac{\dd}{\dd\tau}\left(y^2 g\left(\frac{\omega\zeta}{y^3}\right)\right) =
\dot{y}y g\left(\frac{\omega\zeta}{y^3}\right) - 
\frac{3}{2}\frac{\omega\zeta}{y^2}\dot{y}g^\prime\left(\frac{\omega\zeta}{y^3}\right),
\end{equation}
which allows the reduction of the second-order differential equation for $y$ to a first-order equation,
\begin{equation}
\frac{\dd}{\dd\tau}\dot{y}^2 = \frac{\dd}{\dd\tau}\left(y^2 g\left(\frac{\omega\zeta}{y^3}\right)\right).
\end{equation}
Subsequent integration yields
\begin{equation}
\dot{y}^2 = y^2 g\left(\frac{\omega\zeta}{y^3}\right) + c,
\label{eqn_ydot}
\end{equation}
where the integration constant is determined by requiring $\dot{y}=0$ at turn around $y=1$:
\begin{equation}
c = -g(\omega\zeta),
\end{equation}
yielding the final result for the differential equation defining $y(\tau)$:
\begin{equation}
\dot{y} = \sqrt{y^2g\left(\frac{\omega\zeta}{y^3}\right)-g(\omega\zeta)},
\end{equation}
For small radii $y$ at early times $\tau$, the differential equation can be approximated by
\begin{equation}
\dd\tau \simeq \sqrt{\frac{y}{\omega\zeta}}\left(1+\frac{g(\omega\zeta)}{\omega\zeta}\:\frac{y}{2}\right)\dd y,
\end{equation}
due to $g(\xi)\simeq\xi$ at early times, and can be integrated to yield
\begin{equation}
\tau = \frac{1}{\sqrt{\omega\zeta}}\frac{2}{3}y^\frac{3}{2}
\left(1 + \frac{3}{10}\frac{g(\omega\zeta)}{\omega\zeta}\:y\right).
\label{eqn_ysolution}
\end{equation}
Comparing eqn.~(\ref{eqn_ysolution}) with eqn.~(\ref{eqn_xsolution}) at early times and eliminating $\tau$ yields a solution for $y(x)$, which can be used for defining the time evolution $\Delta(x)$ of the density $\zeta$, restricted to early times:
\begin{equation}
\Delta(x) \equiv \zeta\frac{x^3}{y^3} 
= 1 + \frac{3}{5}\frac{g(\omega\zeta)}{\omega\zeta} \:y.
\end{equation}
The linear density contrast $\delta_c$ is related to the overdensity $\Delta-1$ by computing the density resulting from linear growth,
\begin{equation}
\delta_c = D_+(x_c)\lim_{x\to 0}\left(\frac{\Delta(x)-1}{D_+(x)}\right).
\end{equation}
The overdensity $\delta_c$ is shown in Fig.~\ref{fig_deltac} as a function of redshift $z$, along with the constant value $\delta_c=1.686$ for spherical collapse in a SCDM cosmology. $\delta_c(z)$ falls significantly below the SCDM value at low redshifts with increasing $\alpha$, reaching values as low $\delta_c=1.46$ for modified gravity. Interestingly, the range of $\delta_c$-values corresponds to those obtained in early dark energy models, as proposed by \citet{2006A&A...454...27B} and \citet{2007MNRAS.tmp..641S}, albeit at lower redshifts. Consequently, the observational signature of lowered $\delta_c$ in modified gravity would be an increased abundance of clusters at the relevant redshifts, which will be probed especially by future Sunyaev-Zel'dovich surveys.

The dependence of $\delta_c$ on redshift can be well fitted with the phenomenological function $\delta_c(z) = a - b\exp(-cz)$. Table~\ref{table_fit_params} summarises the values of the parameter $a$, $b$ and $c$, which result from a fit to numerically obtained values for $\delta_c(z)$ in the redshift range $0\leq z\leq 5$, for the range of modified gravity models (characterised by $\alpha$) considered here.

\begin{table}\vspace{-0.1cm}
\begin{center}
\begin{tabular}{lccc}
\hline\hline
$\alpha$	& $a$ 		& $b$		& $c$ 		\\
\hline
$\alpha=0$	& 1.6823	& 0.0129	& 1.5648	\\
$\alpha=0.2$	& 1.6820	& 0.0314	& 1.0912	\\
$\alpha=0.4$	& 1.6806	& 0.0582	& 0.9994	\\
$\alpha=0.6$	& 1.6774	& 0.0942	& 0.9369	\\
$\alpha=0.8$	& 1.6710	& 0.1397	& 0.8722	\\
$\alpha=1$	& 1.6585	& 0.1945	& 0.8020	\\
\hline
\end{tabular}
\end{center}
\caption{Fitting values for the empirical model $\delta_c(z) = a - b\exp(-cz)$ for the range of the modified gravity parameter $\alpha$ considered here, which reproduces the numerical values for $\delta_c(z)$ to sub-percent accuracy in the redshift range $0\leq z\leq 5$.}
\label{table_fit_params}
\end{table}

In the numerical solution, the sequence of steps is as follows: From the specified collapse scale factor $a_c$, the corresponding cosmic time $t_c$ is computed using $t=\int\dd a/(aH)$ with the Hubble function $H(a)$ obtained numerically from the Friedmann equation eqn.~(\ref{eqn_friedmann}). Due to the time-reversal symmetry of eqn.~(\ref{eqn_ydotdot}), turn around takes place at $t_a=t_c/2$, from which $a_a$ can be retrieved. The scale factor at turn around $a_a$ is used to define $\omega$ and $\lambda$ using eqn.~(\ref{eqn_omegam}). $\zeta$ is adjusted numerically by evolving eqn.~(\ref{eqn_ydot}) to the time $\tau^\prime$ at which $\dot{y}(\tau^\prime)=0$, and applying a numerical root finding scheme on $\tau^\prime - \tau(x=1)$, where the dimensionless time $\tau(x=1)$ at turn around $x=1$ results from eqn.~(\ref{eqn_xcollapse}) for given density parameters $\omega$ and $\lambda$. From these results, $\Delta(x)$ and $\delta_c(z)$ follow directly. In the numerical treatment, we use a very good approximation to the implicitly defined Hubble function $H(a)$, which is described in detail in Appendix~\ref{sect_appendix_hubble_approx}.

It is worth noting that $\delta_c$ depends on the matter density $\Omega_m$ from the determination of $\zeta$ via the turn around point in the evolution of the $y$-differential equation, as well as from the growth function $D_+$. In the remainder of the paper, we neglect the influence of $\Omega_m$ on $\delta_c$, as it is of minor influence regarding e.g. cluster number counts compared to first order dependence of the mass function on $\Omega_m$.

% --- section: cluster number counts --- %
\section{cluster number counts and merger rates}\label{sect_number_counts}
Fluctuations in the cosmic density field on the scale $k$ are described by the power spectrum $P(k)$, which results from ensemble averaging of Fourier modes $\delta(\bmath{k})$, $\bra\delta(\bmath{k})\delta(\bmath{k}^\prime)\ket=(2\pi)^3\delta_D(\bmath{k}+\bmath{k}^\prime) P(k)$. The shape of the power spectrum $P(k) \propto k^{n_s} T^2(k)$ is well approximated by the transfer function suggested by \citet{1986ApJ...304...15B} 
\begin{equation}
T(q) = \frac{\ln(1+2.34q)}{2.34q}\left(1+3.89q+(16.1q)^2+(5.46q)^3+(6.71q)^4\right)^{-\frac{1}{4}}
\end{equation}
where the the wave vector $k=q\Gamma$ is rescaled with the with shape parameter $\Gamma \simeq \Omega_m h$. The fluctuation amplitude is normalised to the value $\sigma_8$ on the scale $R=8~\mathrm{Mpc}/h$ with a spherical top-hat filter.

The number density of clusters forming by spherical collapse in a fluctuating density field is given by 
\begin{equation}
n(M,z)\dd M = 
\sqrt{\frac{2}{\pi}}\: \rho_0\: \Delta(z,M)
\frac{\dd\ln\sigma(M)}{\dd\ln M}
\exp\left(-\frac{\Delta^2(z,M)}{2}\right)
\frac{\dd M}{M^2}
\end{equation}
\citep{1974ApJ...187..425P, 1991ApJ...379..440B} with background density $\rho_0=\Omega_m\rho_\mathrm{crit}$, where
\begin{equation}
\Delta(z,M) \equiv \frac{\delta_c(z)}{D_+(z)\sigma(M)}
\end{equation}
abbreviates the quotient between the linear overdensity $\delta_c(z)$ needed for collapse at $z$, the growth function $D_+(z)$ and the variance $\sigma(M)$ of the density field on the mass scale $M$, which is related to the spatial filtering scale $R$ by $M=4\pi/3 \rho_\mathrm{crit}\Omega_M R^3$. The Press-Schechter mass function is used rather than the extension by Sheth and Tormen because the latter describes ellipsoidal collapse, where the modified gravity collapse equations might bot be applicable.

The logarithmic derivative of the fluctuation amplitude with logarithmic mass is reformulated as a derivative with respect to spatial scale $R$:
\begin{equation}
\frac{\dd\ln\sigma}{\dd\ln M} = \frac{1}{3}\frac{\dd\ln\sigma}{\dd\ln R} = \frac{R}{3\sigma}\frac{\dd\sigma}{\dd R}
=\frac{R^2}{6\sigma^2}\frac{\dd\sigma^2}{\dd R},
\end{equation}
where the fluctuation amplitude $\sigma^2(R)$ and its derivative $\dd\sigma^2/\dd R$ are given by:
\begin{eqnarray}
\sigma^2(R)
& = & \frac{1}{2\pi^2}\int\dd k\:k^2 P(k)\left(\frac{3}{kR}j_1(kR)\right)^2,\label{eqn_sigma_filter}\\
\frac{\dd\sigma^2}{\dd R}
& = & \frac{1}{\pi^2}\int\dd k\:k^3 P(k)\left(\frac{3}{kR}\right)^2 j_1(kR)j_2(kR).
\end{eqnarray}
$j_\ell(x)$ denotes the spherical Bessel function of the first kind of order $\ell$ \citep{1972hmf..book.....A}.

\begin{figure}
\resizebox{\hsize}{!}{\includegraphics{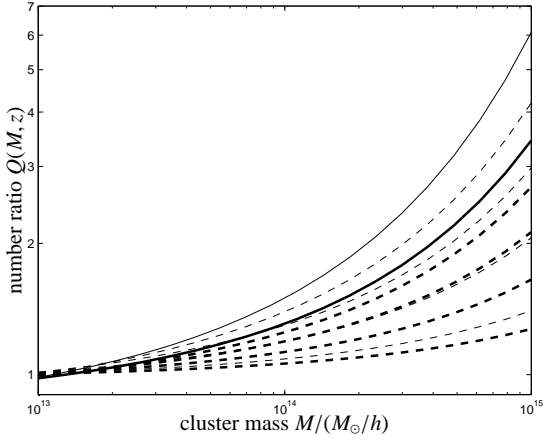}}
\caption{Ratio $Q(M,z)$ of the mass functions $q(M,z)$ relative to $\Lambda$CDM ($\alpha=0$), at redshifts $z=0.3$ (thick lines) and $z=0.7$ (thin lines), for values of $\alpha$ from $\alpha=0.2$ in steps of $\Delta\alpha=0.2$ (dashed) up to $\alpha=1$ (modified gravity, solid).}
\label{fig_cdm_mass}
\end{figure}

Multiplication of $n(M,z)$ with the evolution $\dd V/\dd z$ of the comoving volume with redshift yields the cluster density per unit redshift,
\begin{equation}
q(M,z) = n(M,z)\:\frac{\dd V}{\dd z}.
\end{equation}
The quantity $Q(M,z)$, which is defined as the ratio of $q(M,z)$ for a modified gravity model characterised with $\alpha$ relative to $\Lambda$CDM,
\begin{equation}
Q(M,z) = \frac{q_\alpha(M,z)}{q_{\alpha=0}(M,z)}
\end{equation}
is plotted in Fig.~\ref{fig_cdm_mass} for redshifts $z=0.3$ and $z=0.7$, where the PLANCK and SPT surveys are expected to yield most of their detections. 

Modified gravity significantly enhances the cluster number counts which is mainly caused by the lower values of the linear overdensity for spherical collapse $\delta_c$ at low redshifts. At a redshift of $z=0.3$, corresponding to the peak in the redshift distribution of PLANCK's detections, on can expect to detect almost 50\% more clusters at masses of $10^{14}M_\odot/h$. This effect is more pronounced at increased redshift: SPT is yield most of the detections at a redshift of $z=0.7$, where number counts of $10^{14}M_\odot/h$-mass objects would be almost twice as large in modified gravity models compared to $\Lambda$CDM. At higher masses of $10^{15}M_\odot/h$, the increase in cluster number density amounts to factors of three (at $z=0.3$) up to six (at z=0.7), but the small spatial density of very massive objects at high redshift will make this difficult to observe. In contrast, the difference between modified gravity cosmologies and $\Lambda$CDM vanishes at small masses, because low-mass objects form early the matter-dominated phase, where $\delta_c$ has not deviated significantly from its canonical value of $1.686$. The enhancement in cluster number density depends exponentially on the modified gravity parameter $\alpha$.

\begin{figure}
\resizebox{\hsize}{!}{\includegraphics{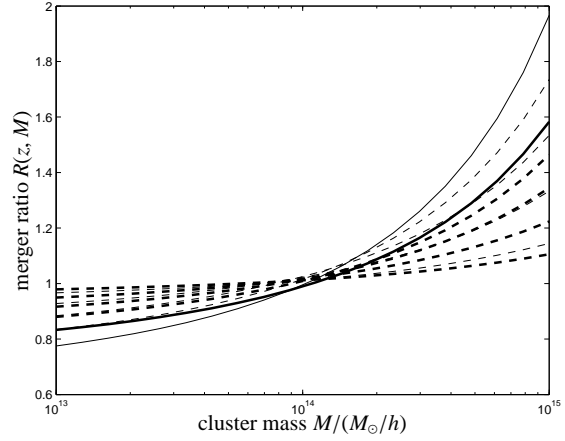}}
\caption{Ratio $R(M,z)$ of the merger densities $r(m,z)$ relative to $\Lambda$CDM ($\alpha=0$), at redshifts $z=0.3$ (thick lines) and $z=0.7$ (thin lines), for values of $\alpha$ from $\alpha=0.2$ in steps of $\Delta\alpha=0.2$ (dashed) up to $\alpha=1$ (modified gravity, solid). The mass ratio has been set to $M_2/M_1=2$.}
\label{fig_cdm_merger}
\end{figure}

The density $m(M,z)$ of merging events as a function of redshift and mass ratio $M_2/M_1$ can be determined by extending the Press-Schechter formalism outlined above \citep{1991ApJ...379..440B, 1993MNRAS.262..627L, 1994MNRAS.271..676L} for an expression for the merger probability as a function of mass, mass difference and cosmic time,
\begin{equation}
\frac{\dd^2p(M,z)}{\dd\ln\Delta M\:\dd t}.
\label{eqn_merger_density}
\end{equation}
For the purpose of this work, the merger density is multiplied with the volume element $\dd V/\dd z$ and the evolution $\dd t/\dd z$ of cosmic time with redshift in order to yield the merger density per unit redshift,
\begin{equation}
r(M,z) = 
\left(n(M,z)\: \frac{\dd V}{\dd z}\right) \: 
\left(\frac{\dd^2p(M,z)}{\dd\ln\Delta M\:\dd t}\:\frac{\dd t}{\dd z}\right)
\end{equation}
$R(M,z)$ is defined as the ratio of $r(M,z)$ between a modified gravity model and $\Lambda$CDM,
\begin{equation}
R(M,z) = \frac{r_\alpha(M,z)}{r_{\alpha=0}(M,z)}.
\end{equation}
The quantity $R(M,z)$ is plotted in Fig.~\ref{fig_cdm_merger} for redshifts $z=0.3$ and $z=0.7$ for a fixed merger mass ratio of $M_2/M_1=2$. Merger densities exhibit a similar behaviour at high masses of $10^{15}M_\odot/h$ compared to the mass function, where the modified gravity model predics numbers roughly factors of 1.6 (PLANCK) and two (SPT) times higher at the relevant redshifts. Interestingly, the merger density is decreased in modified gravity models relative to $\Lambda$CDM at small masses; this decrease is due to the unique weighting of the merger density with the evolution of cosmic time with redshift $\dd t/dz$, the evolution of comoving volume $\dd V/\dd z$ and the Hubble function $H(a)$, resulting from the conversion of a time derivative of $\delta_c/D_+$ in eqn.~(\ref{eqn_merger_density}) to a derivative in scale factor.

% --- section: SZ surveys --- %
\section{Constraints from future SZ surveys}\label{sect_szsurvey}
In this section, constraints on cosmological parameters including the modified gravity parameterisation $\alpha$ from future high-yield Sunyaev-Zel'dovich cluster surveys are estimated in a Fisher-matrix analysis. Following \citet{2004ApJ...613...41M}, a measurement of the cluster number density per unit redshift interval, observed in the solid angle $\Delta\Omega$,
\begin{equation}
N(z) \equiv \frac{\dd n}{\dd z} = \frac{\Delta\Omega}{4\pi}\: \frac{\dd V}{\dd z} \: \int_{M_\mathrm{min}(z)}^\infty\dd M\: n(M,z),
\label{eqn_cluster_abundance}
\end{equation}
is considered with a simple flux threshold $S_\mathrm{min}$, expressed as a minimal mass $M_\mathrm{min}$ dependent on redshift. Using the formula given by \citet{2004ApJ...613...41M}, the flux distortion $\Delta S(\nu)$ caused by the SZ-effect (in $m$Jansky) at frequency $\nu$ is related to the angular diameter distance $d_A(z)$ (in $\mathrm{Mpc}$) and the cluster mass $M$ (in $M_\odot$) by
\begin{equation}
\Delta S(\nu) = 
\frac{1}{d_A^2(z)}f_\mathrm{ICM}\: g_\mathrm{SZ}(\nu)\: A_\mathrm{SZ}\: H^\frac{2}{3}(z)\: M^{1+\beta}\: (1+z),
\end{equation}
where $f_\mathrm{ICM}=0.12$ and $\log A= -16.71$ \citep{2001MNRAS.325.1533D,2001A&A...368..749F}. For the slope of the temperature-mass relation, $T(M)\propto M^\beta$, the value $\beta=0.65$ is chosen \citep{1999ApJ...517..627M}. The frequency dependence of the Sunyaev-Zel'dovich effect \citep{1972CoASP...4..173S, 1980ARA&A..18..537S} is given by
\begin{equation}
g(x) =  \frac{x^4\exp(x)}{(\exp(x)-1)^2}\left(x\frac{\exp(x)+1}{\exp(x)-1}\right)
\end{equation}
with the dimensionless frequency $x=h\nu/(k_B T_\mathrm{CMB})$. In addition to the flux criterion, a lower mass threshold of $10^{14}M_\odot/h$ is imposed, whichever yields the larger limiting mass.

From the cluster number counts $N_i$ in a redshift bin centered at $z_i$ and the derivative $\partial N_i/\partial p_\mu$ with respect to the parameter $p_\mu$ the Fisher matrix $F_{\mu\nu}^\mathrm{SZ}$ can be constructed \citep{2001ApJ...560L.111H}:
\begin{equation}
F_{\mu\nu}^\mathrm{SZ} = 
\sum_i \frac{\partial N_i}{\partial p_\mu} \: \frac{1}{N_i} \: \frac{\partial N_i}{\partial p_\nu},
\end{equation}
assuming a Gaussian likelihood and statistically independent number counts $N_i$ in each redshift bin.The summation extends over the redshift range from $z_\mathrm{min}=0.01$ to $z_\mathrm{max}=2.0$ in steps of $\Delta z = 0.02$. 

Using the Cram{\'e}r-Rao inequality, the accuracy on a single parameter $\Delta p_\mu$ is equal to $\sqrt{(F^{-1})_{\mu\mu}}$. $\chi^2$ for a pair of parameters $(p_\mu,p_\nu)$ can be computed from the inverse $(F^{-1})_{\mu\nu}$ of the Fisher matrix,
\begin{equation}
\chi^2 = 
\left(
\begin{array}{c}
\Delta p_\mu \\
\Delta p_\nu
\end{array}
\right)^t
\left(
\begin{array}{cc}
(F^{-1})_{\mu\mu} & (F^{-1})_{\mu\nu} \\
(F^{-1})_{\nu\mu} & (F^{-1})_{\nu\nu}
\end{array}
\right)^{-1}
\left(
\begin{array}{c}
\Delta p_\mu \\
\Delta p_\nu
\end{array}
\right),
\end{equation}
where $\Delta p_\mu = p_\mu - p_\mu^{\Lambda CDM}$. For parameter pairs the isoprobability contour at
\begin{equation}
\Delta\chi^2_n = -\ln\:\mathrm{erfc}^2\left(\frac{n}{\sqrt{2}}\right),
\end{equation}
encloses a fraction of $\mathrm{erf}(n/\sqrt{2})$ for of the admissible parameter space, corresponding to a confidence level of $n\sigma$. $\mathrm{erfc}(x)=1-\mathrm{erf}(x)$ is the complementary error function \citep{1972hmf..book.....A}. 

For the purpose of this paper, the PLANCK and SPT Sunyaev-Zel'dovich surveys are considered, which are expected to yield cluster catalogues comprising $10^4$ and $3\times10^4$ entries, respectively \citep[c.f.][]{2005A&A...429..417M,2006astro.ph.11567G, 2007MNRAS.377..253M}. The principal characteristics of these surveys are summarised in table~\ref{table_sz_survey}. As cosmological parameters, $\Omega_m$ (assuming spatial flatness), $\sigma_8$ and $n_s$ are considered, alongside the modified gravity parameter $\alpha$ and the slope of the temperature-mass relation $\beta$. In general, the increased spatial number density of clusters caused by deviations in the gravitational law has to compete against a slower evolution of the comoving volume with redshift $\dd V/\dd z$, since the two terms in eqn.~(\ref{eqn_cluster_abundance}) are affected in opposite ways. 

\begin{table}\vspace{-0.1cm}
\begin{center}
\begin{tabular}{lll}
\hline\hline
		& PLANCK 			& SPT 				\\
\hline
frequency 	& $\nu=353$~GHz 		& $\nu=150$~GHz			\\
flux		& $S_\mathrm{min}=100$~$m$Jy 	& $S_\mathrm{min}=10$~$m$Jy	\\
solid angle 	& $\Delta\Omega=4\pi$ 		& $\Delta\Omega=4\pi/10$	\\
\hline
\end{tabular}
\end{center}
\caption{Frequency $\nu$, flux threshold $S_\mathrm{min}$ and the solid angle $\Delta\Omega$ of the Sunyaev-Zel'dovich surveys to be carried out by PLANCK and by SPT, respectively.}
\label{table_sz_survey}
\end{table}

\begin{figure*}
\resizebox{\hsize}{!}{\includegraphics{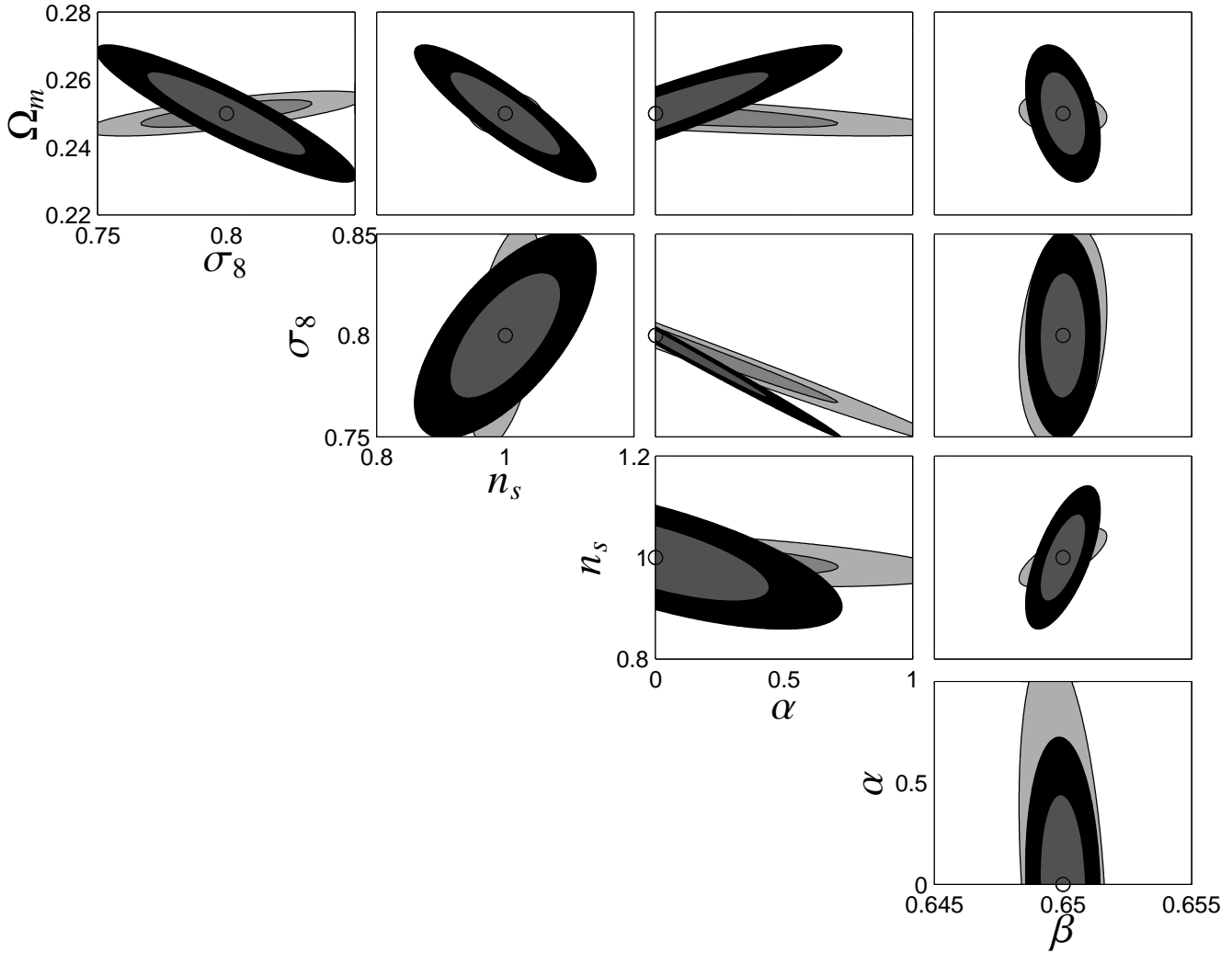}}
\caption{Forecasts for parameter degeneracies for the PLANCK (light gray) and SPT (dark gray) Sunyaev-Zel'dovich surveys in a Fisher-matrix analysis. The fiducial $\Lambda$CDM cosmology is marked by a circle: $\Omega_m = 0.25$ (assuming spatial flatness), $\sigma_8=0.8$, $n_s=1.0$, the modified gravity parameterisation $\alpha=0$ and the slope of the temperature-mass relation $\beta=0.65$. The ellipses correspond to $1\sigma$ and $2\sigma$ confidence intervals.}
\label{fig_fisher}
\end{figure*}

The results from the Fisher matrix analysis are compiled in Fig.~\ref{fig_fisher}. The SZ-observations alone yield accuracies of a few percent on the parameters $\Omega_m$ and $\sigma_8$, and marginally worse constraints on $n_s$. PLANCK and SPT would be able to exclude values of $\alpha>0.4$, which compare favourably to contemporary constraints from SNIa observations \citep[e.g. ][ who find the entire range of $\alpha$ consistent with observations]{2006PhLB..642..432F}. SPT's smaller power in constraining $\alpha$ is due to the fact that its SZ-sample contains more high-redshift clusters, which have formed in an epoch close to matter domination where the influence of modified gravity is smaller. An important consistency check is that the accuracy of measuring $\alpha$ derived from the SZ-surveys compares well to the allowed range of values for the dark energy equation of state $w(a)$ \citep{2004ApJ...613...41M} of $\Delta w=0.1$ (c.f. Fig.~\ref{fig_eos}, where the $w(a)$ for a dark energy model with the identical expansion history as the modified gravity model is derived).

The $\alpha$-$\sigma_8$ degeneracy is readily explained by the fact that clusters form more easily in modified gravity, which has to be compensated by lowering the value of $\sigma_8$. Tilting the CMB spectrum and using smaller values for $n_s$ yields more massive clusters if the normalisation $\sigma_8$ is kept fixed. Whether this is observable is a matter of the redshift distribution of the experiment: At low redshifts of $z\simeq0.3$, where most of the PLANCK-detections will be situated, the opposite effects of increasing $\alpha$ on the cluster density and on the volume element almost cancel each other, whereas at high redshifts of $z\simeq0.7$, corresponding to the redshift at which SPT will find most of the objects, the decrease in $\dd V/\dd z$ caused by modified gravity dominates.

In the $\alpha$-$\Omega_m$ degeneracy both parameters impact on the cluster density and the evolution of the comoving volume element: At low redshifts, the increased number densities of clusters due to modified gravity is balanced by low choices of $\Omega_m$, which affect both the cluster number density as well as the volume element. At higher redshifts, however, the volume element becomes the dominating factor as it is decreased by modified gravity, only balanced by larger comoving densities due to higher choices of $\Omega_m$.

A crucial point is the understanding of flux threshold due to the strong dependence of the observed number of clusters on the particular value of the limiting minimal mass as a function of redshift. The key parameter is the slope of the temperature-mass relation $\beta$, which has to be known to the level of a few percent, and which would be controlled by self-calibration techniques \citep[e.g.][]{2004ApJ...613...41M} in a realistic application.

\begin{figure*}
\begin{tabular}{cc}
\resizebox{0.47\hsize}{!}{\includegraphics{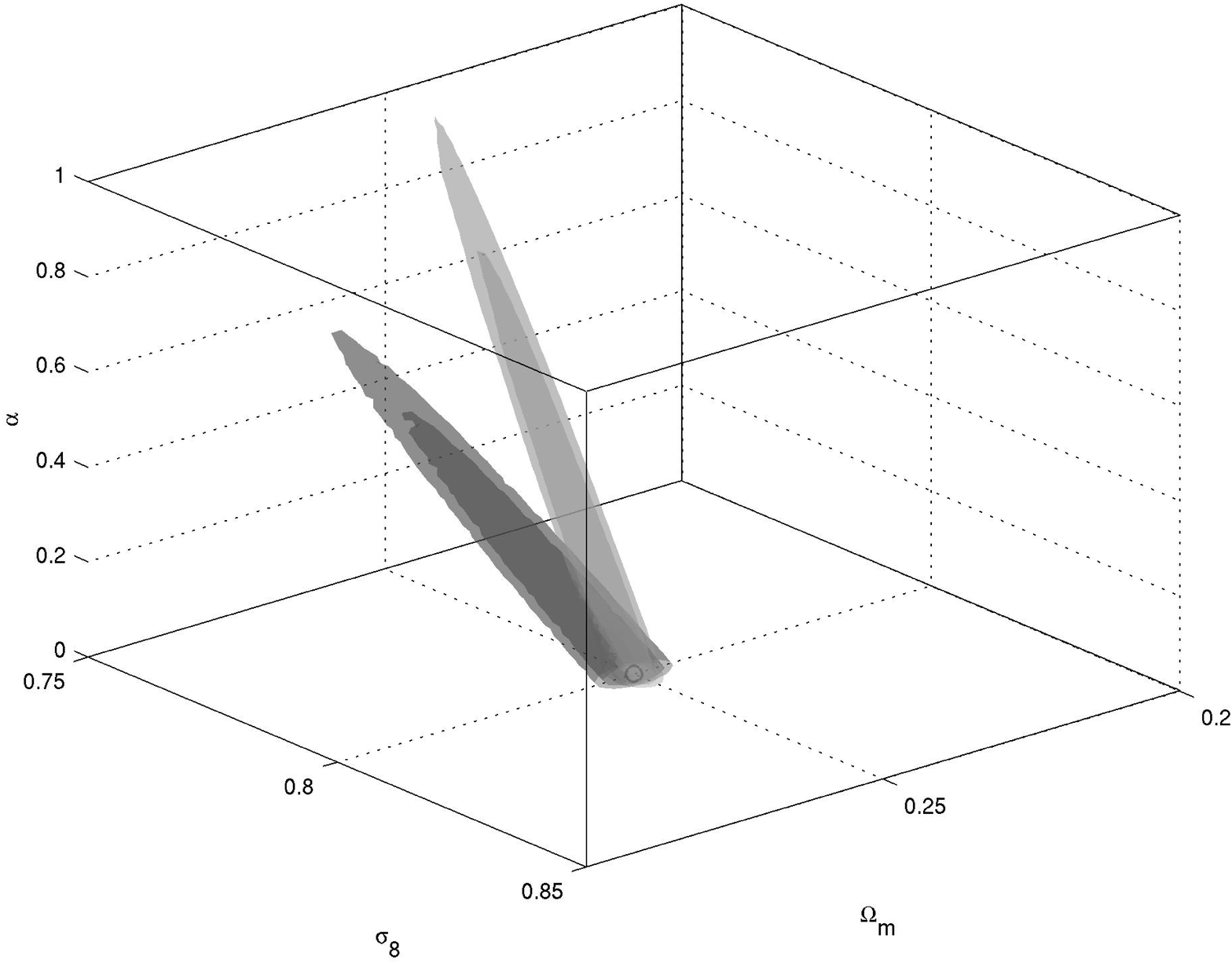}} &
\resizebox{0.47\hsize}{!}{\includegraphics{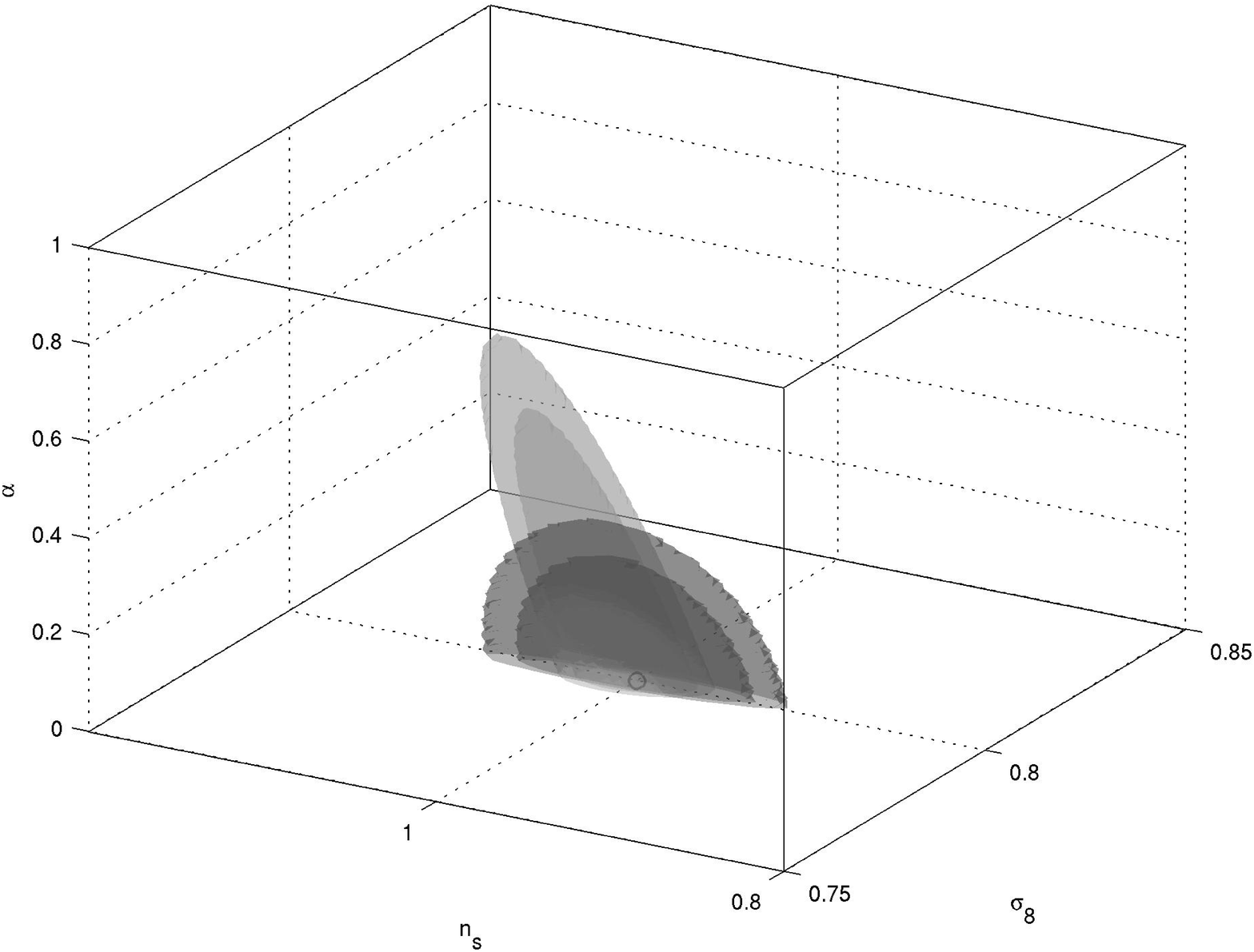}} 
\end{tabular}
\caption{Simultaneous constraints on three cosmological parameters from the PLANCK (light gray) and SPT (dark gray) Sunyaev-Zel'dovich surveys, for the parameters $\Omega_m$, $\sigma_8$ and $\alpha$ (left panel) and $\sigma_8$, $n_s$ and $\alpha$ (right panel). The circles mark the parameter choices $\Omega_m=0.25$, $\sigma_8=0.8$, $n_s=1$ and $\alpha=0$ for the fiducial $\Lambda$CDM cosmology. The nested ellipsoid surfaces correspond to $1\sigma$ and $2\sigma$ confidence intervals.}
\label{fig_fisher3d}
\end{figure*}

Additionally, Fig.~\ref{fig_fisher3d} shows three-dimensional degeneracy ellipsoids between a pair of cosmological parameters and the modified gravity parameter $\alpha$. It emphasises the fact that a strong prior on the cosmological parameters relevant for cluster formation can significantly improve the constraint on modified gravity. Especially SPT should be able to yield a constraint on $\alpha$ with independent measurments of $\Omega_m$, $n_s$ and especially $\sigma_8$.

Strong prior constraints on $\Omega_m$, $\sigma_8$ and $n_s$ are given by the observation of e.g. CMB temperature and polarisation anisotropies: For independent observational channels, the combined likelihood of the parameters is the product of the individual likelihoods, and hence
\begin{equation}
F_{\mu\nu} = F_{\mu\nu}^\mathrm{SZ} + F_{\mu\nu}^\mathrm{CMB},
\end{equation}
where $F_{\mu\nu}^\mathrm{CMB}$ is derived for PLANCK's observation of temperature anisotropy spectrum $C_T(\ell)$, the polarisation anisotropy spectrum $C_E(\ell)$ and the cross-spectrum $C_C(\ell)$. The construction of the Fisher matrix $F_{\mu\nu}^\mathrm{CMB}$ and details such as the assumed noised levels and beam dimensions are given in Appendix~\ref{sect_appendix_cmb}. 

Expected errors on individual cosmological parameters with and without CMB priors are given in table~\ref{table_fisher}. Apart from percent accuracy on the cosmological parameters $\Omega_m$, $\sigma_8$ and  $n_s$, the minimum variance bound on the modification of gravity derived from the Fisher-matrices amounts to 0.46 from PLANCK and to 0.29 from SPT, i.e. from a cluster survey alone one should be able to exclude modified gravity models (corresponding to $\alpha=1$) with a statistical significance of $3\sigma$. The improved measurement is able to constrain the modified gravity parameter $\alpha$ significantly better to values below $\Delta\alpha=0.063$ and $\Delta\alpha=0.035$, for PLANCK and SPT, respectively.

\begin{table}\vspace{-0.1cm}
\begin{center}
\begin{tabular}{llll}
\hline\hline
PLANCK 			& SPT 				& PLANCK $+$ CMB		& SPT $+$ CMB	\\
\hline
$\Delta\Omega_m=0.027$ 	&$\Delta\Omega_m=0.008$		& $\Delta\Omega_m=0.002$ 	&$\Delta\Omega_m=0.001$\\
$\Delta\sigma_8=0.082$ 	&$\Delta\sigma_8=0.021$		& $\Delta\sigma_8=0.002$ 	&$\Delta\sigma_8=0.002$\\
$\Delta n_s=0.023$ 	&$\Delta n_s=0.057$		& $\Delta n_s=0.002$ 		&$\Delta n_s=0.002$\\
$\Delta\alpha=0.469$ 	&$\Delta\alpha=0.292$		& $\Delta\alpha=0.063$ 		&$\Delta\alpha=0.035$\\
\hline
\end{tabular}
\end{center}
\caption{Fisher-matrix forecasts for parameter accuracies ($1\sigma$ confidence intervals) from the PLANCK and SPT Sunyaev-Zel'dovich surveys (first and second column), and the corresponding accuracies after adding prior constraints on $\Omega_m$, $\sigma_8$ and $n_s$ from PLANCK's CMB temperature and polarisation spectra (third and fourth column). $\Lambda$CDM is the fiducial cosmology, with parameters $\Omega_m=0.25$, $\sigma_8=0.8$, $n_s=1$ and $\alpha=0$.}
\label{table_fisher}
\end{table}

These results are quite comparable to the findings of \citet{2006astro.ph..9028T}, who combined supernovae and cluster counts for constraining DGP-gravity, which is achievable with a few hundred clusters from an SZ-observation, given a prior tight constraint on $\sigma_8$. In their analysis, the influence of modified gravity on the Hubble expansion and on the growth equation was considered, and the density threshold by spherical collapse was approximated to be constant. \citet{2006astro.ph..9028T} point out that a pure geometrical measurement can never distinguish between modified gravity and a suitably tuned dark energy model, and that additional information such as a measurement of the growth function is needed.

% --- section: Summary --- %
\section{Summary}\label{sect_summary}
In this paper, the influence of modified gravity on Friedmann dynamics, the homogeneous growth of the cosmic density field and the formation of galaxy clusters is considered. The alteration of the gravitational law is motivated by DGP-gravity, and introduced as an additional term in the Friedmann equations, which are consistently applied to the background dynamics as well as for the formation of individual clusters. A parameter $\alpha$ allows interpolation between the two limiting cases: gravitation with a cosmological constant $\Lambda$ and DGP-gravity.
\begin{itemize}
\item{The Friedmann dynamics of the universe is remarkably little affected by the modification of gravity, as the Hubble function $H(a)$ for our range of models differs by at most 10\% at a scale factor of $a=0.5$. This translates to comparable variations on the evolution $\dd V/\dd z$ of volume with redshift and the evolution $\dd t/\dd z$ of cosmic time with redshift.}

\item{The growth function shows significant suppression of growth at early times for modified gravity, very similar to dark energy models with evolving equation of state, where the first derivative of the equation of state parameter $w(a)$ is positive at zero redshift. The largest spread between the models considered here occurs around $a=0.3$. The changed behaviour of the growth function is both caused by the decrease in matter density $\Omega_m(a)$ as well as by the scaling of the Hubble-function with scale factor.}

\item{The spherical collapse equations can be solved numerically for the modified gravity models motivated by DGP, resulting in the linear overdensity for spherical collapse $\delta_c$ as a function of collapse redshift $z$. One notices a significant decrease in $\delta_c$ by almost 20\% compared to SCDM. The range of values for $\delta_c$ is quite comparable to those obtained in collapse with early dark energy. An empirical fitting formula of the shape $\delta_c(z)=a-b\exp(-cz)$ was found to yield a good approximation.}

\item{Cluster number counts are significantly enhanced by modified gravity, due to the lower values of the linear overdensity for spherical collapse $\delta_c$ at low redshifts. At a redshift of $z=0.3$, where the majority of PLANCK SZ-clusters will be situated, as much as 50\% more clusters with masses of $10^{14}M_\odot/h$ have formed, whereas at a redshift of $z=0.7$, corresponding to the peak in the cluster redshift distribution forecast for SPT, number counts might be 70\% larger. At higher masses of $10^{15}M_\odot/h$, the increase in cluster number density amounts to factors of three (at $z=0.3$) up to six (at z=0.7), but the sparsity of very massive objects at high redshift will make this difficult to observe. At low masses, on the contrary, there is hardly a difference between DGP-like modified gravity models and $\Lambda$CDM, which is due to the fact that these objects form early well in the matter-dominated phase. The enhancement in cluster number density depends exponentially on the modified gravity parameter $\alpha$.}

\item{Merger densities exhibit a similar behaviour at high masses of $10^{15}M_\odot/h$, where the model predics an increase by factors of roughly up to two at the redshifts considered. At low masses, however, the merger density falls below the prediction of $\Lambda$CDM, which is carried by the unique weighting of the merger density with the evolution of cosmic time with redshift $\dd t/dz$, the evolution of comoving volume $\dd V/\dd z$ and the Hubble function $H(a)$.}

\item{The impact of modified gravity on cluster formation is most notable at small redshifts below unity, making it a target for Sunyaev-Zel'dovich cluster surveys. In this paper the cluster yield of the SZ-surveys to be carried out by PLANCK and SPT was modelled with a simple flux threshold as the detection criterion. The Fisher-matrix analysis with   $\Lambda$CDM as the fiducial cosmological model found that on themselves, the SZ-surveys alone are able to constrain the modified gravity parameter $\alpha$ with an accuracy of $\Delta\alpha\simeq0.47$, albeit with strong degeneracies. Adding in CMB constraints on $\Omega_m$, $\sigma_8$ and $n_S$ significantly lowered the uncertainty to $\Delta\alpha=0.04$, making the combined SZ- and CMB-observations a powerful observational probe.}
\end{itemize}

As last points is it worth noting that the above described model for phenomenological DGP-gravity has the same number of parameters as $\Lambda$CDM and would not be disfavoured on grounds on Bayesian model selection. The feature that cluster formation takes place at higher redshifts compared to $\Lambda$CDM might have important implications on the arclet problem in gravitational lensing \citep{2005A&A...442..413M} and on the CBI-anomaly \citep{2001ApJ...549L...1P, 2005ApJ...626...12B} as it naturally causes a higher level of Sunyaev-Zel'dovich background due to high-redshift clusters with a low contemporary value of $\sigma_8$.

% --- section: Acknowledgements --- %
\section*{Acknowledgements}
We would like to thank Rob Crittenden for valuable comments.

% --- section: bibliography --- %
\bibliography{bibtex/aamnem,bibtex/references}
\bibliographystyle{mn2e}

% --- appendix --- %
\appendix

% --- appendix: equation of state --- %
\section{Equation of state}\label{sect_appendix_eos}
\begin{figure}
\resizebox{\hsize}{!}{\includegraphics{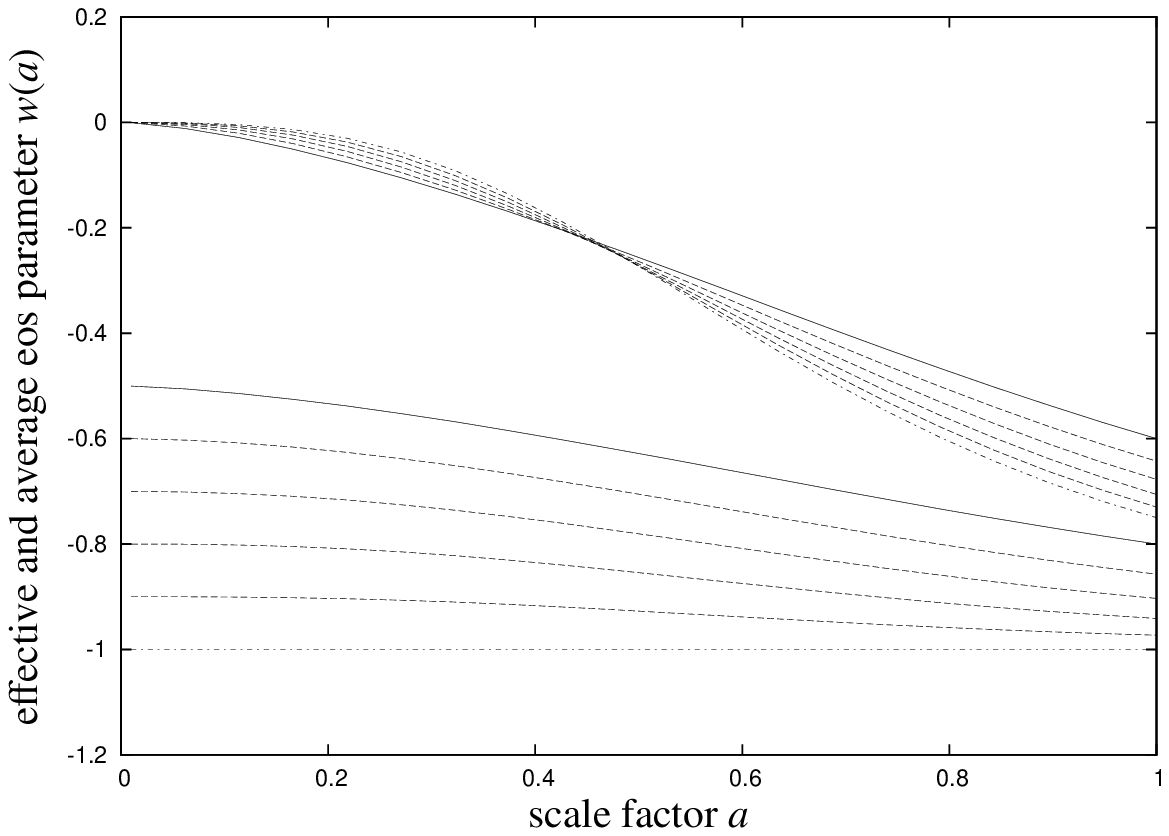}}
\caption{Average equation-of-state parameter $w_\mathrm{avg}(a)$ (equal to 0 at early times) and effective equation-of-state parameter $w_\mathrm{eff}(a)$ as a function of scale factor $a$, for values of $\alpha=0$ (corresponding to $\Lambda$CDM, dash-dotted), $\alpha=1$ (DGP-gravity, solid) and intermediate values in steps of $\Delta\alpha=0.2$ (dashed).}
\label{fig_eos}
\end{figure}

The relation between the modified gravity model and the equation of state $w(a)$ of a dark energy model with identical expansion history is derived in this section. A dark energy fluid with equation of state $w(a)$ gives rise to an additional term $\Delta H^2(a)$ in the Friedmann equation, given by
\begin{equation}
\Delta H^2(a) = \exp\left(3\int_a^1\dd a\:\frac{1+w(a)}{a}\right),
\end{equation}
which can be solved for $w(a)$ for a given Hubble function $H(a)$. By equating $\Delta H^2(a)$ with the modification in the Friedmann equation $(1-\Omega_m)H^\alpha(a)$ the equation of state $w_\mathrm{eff}(a)$ can be derived for a dark energy model with identical expansion history:
\begin{equation}
w_\mathrm{eff}(a) = -\left(1 + \frac{\alpha}{3}\frac{\dd\ln H}{\dd\ln a}\right).
\end{equation}
Another interesting quantity is the equation of state $w_\mathrm{avg}(a)$, averaged over all relevant cosmological fluids. In this case, $\Delta H^2$ has to be equal to the complete Friedmann equation $H^2(a)=(1-\Omega_m)H^\alpha(a)+\Omega_m/a^3$, which results in
\begin{equation}
w_\mathrm{avg}(a) = -\left(1 + \frac{2\alpha}{3}\frac{\dd\ln H}{\dd\ln a}\right).
\end{equation}
Fig.~\ref{fig_eos} compares the average equation of state $w_\mathrm{avg}(a)$ and the effective equation of state $w_\mathrm{eff}(a)$ as a function of scale factor. The effective equation of state $w_\mathrm{eff}(a)$ decreases slowly from -0.5 to -0.8 for the DGP-type models, is constant $w=-1$ for $\Lambda$CDM and smoothly interpolates between these two limiting cases for choices of $\alpha=0{\ldots}1$. On the contrary, the average equation of state $w_\mathrm{avg}(a)$ assumes a value of zero at $a=0$ for all modified gravity models, and decreases towards values in the interval $-0.6{\ldots}-0.8$, with $\Lambda$CDM showing the most abrupt and the DGP-emulating model showing the most gradual transition.

Comparing the results in Fig.~\ref{fig_eos} to current measurements of the equation of state of dark energy, e.g. using weak gravitational lensing \citep{2006ApJ...644...71J}, the cosmic microwave background \citep{2003ApJS..148..195V}, supernovae \citep{2004AIPC..743....3R}, baryon acoustic oscillations \citep{2007arXiv0705.3323P} or combinations of these observational channels \citep{2001PhRvD..64j3508B, 2003PhRvD..68d3509M}, which are usually stated in terms of the parameterisation $w(a)\simeq w_0 + (1-a)w_a$ \citep[introduced by ][]{2003MNRAS.346..573L} shows that these experiments only marginally exclude DGP-gravity and are in agreement with the values of $w_0=-0.9{\ldots}-1.0$ and $w_a=0.3{\ldots}0$ resulting from more conservative choices of $\alpha$.

% --- appendix: growth function --- %
\section{Growth function}\label{sect_appendix_growth}
\begin{figure}
\resizebox{\hsize}{!}{\includegraphics{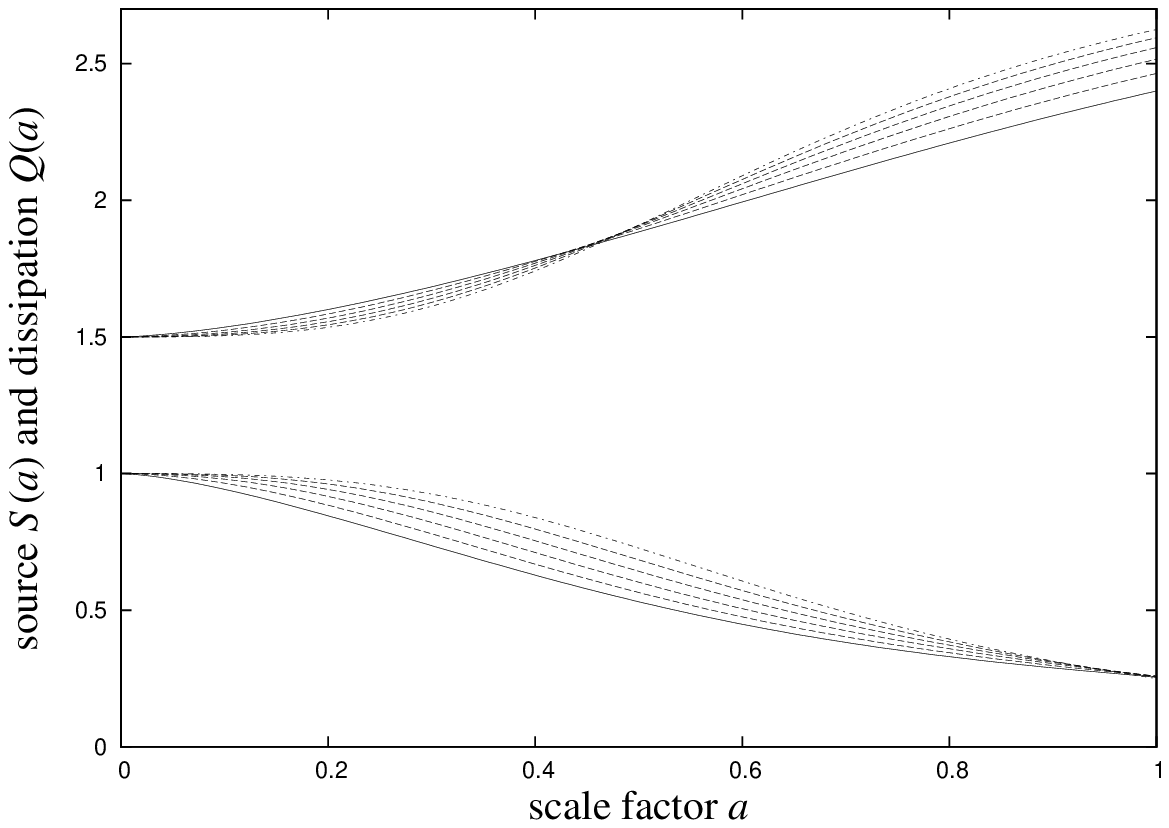}}
\caption{Comparison of the damping term $Q(a)=3+\dd\ln H(a)/\dd\ln a$ (equal to $3/2$ at early times) and the source term $S(a)=G(a)\Omega_m(a)$ (equal to 1 at early times) in the growth equation as a function of scale factor $a$, for values of $\alpha=0$ (corresponding to $\Lambda$CDM, dash-dotted), $\alpha=1$ (DGP-like gravity, solid) and intermediate values in steps of $\Delta\alpha=0.2$ (dashed).}
\label{fig_growth_comp}
\end{figure}

The influence of modified gravity on the growth equation is discussed in this section. Generally, The growth equation eqn.~(\ref{eqn_growth}) has the shape
\begin{equation}
\frac{\dd^2}{\dd a^2}D_+(a) + \frac{Q(a)}{a}\frac{\dd}{\dd a}D_+(a) = \frac{3S(a)}{2a^2}\:D_+(a),
\end{equation}
where a source term $S(a)$ and a dissipation term $Q(a)$ can be identified,
\begin{equation}
Q(a) = 3+\frac{\dd\ln H(a)}{\dd\ln a} ,\quad
S(a) = G(a)\:\Omega_m(a).
\end{equation}
A comparison between the source and damping terms in the growth equation is provided by Fig.~\ref{fig_growth_comp}: Increasing the value of $\alpha$ makes the transition of the damping term $Q(a)$ from $Q(a=0)=\frac{3}{2}$ in the matter dominated epoch to the modified gravity dominated epoch more gradual, due to the influence of modified gravity on the scaling of the Hubble function. 

The source term $S(a)=G(a)\:\Omega_m(a)$ with contributions both from the evolution $\Omega_m$ of the matter density, which is proportional to $H^{-2}(a)$ as well as from modified gravity as a consequence of Birkhoff's theorem \citep{2004PhRvD..69d4005L, 2004PhRvD..69l4015L}, smoothly decreases from $S(a)=1$ at $a=0$ to $S(a)=\Omega_m$ today, and the choice of larger values of $\alpha$ make the source term drop at earlier times. In summary, the suppression of structure formation is caused simultaneously by the decrease in the source term $S(a)$ and the increase in the damping $Q(a)$, while larger choices of $\alpha$ are able to suppress structure formation more effectively at early times.

% --- appendix: approximation to the Hubble function --- %
\section{Approximate solution for the Hubble-function}\label{sect_appendix_hubble_approx}
\begin{figure}
\resizebox{\hsize}{!}{\includegraphics{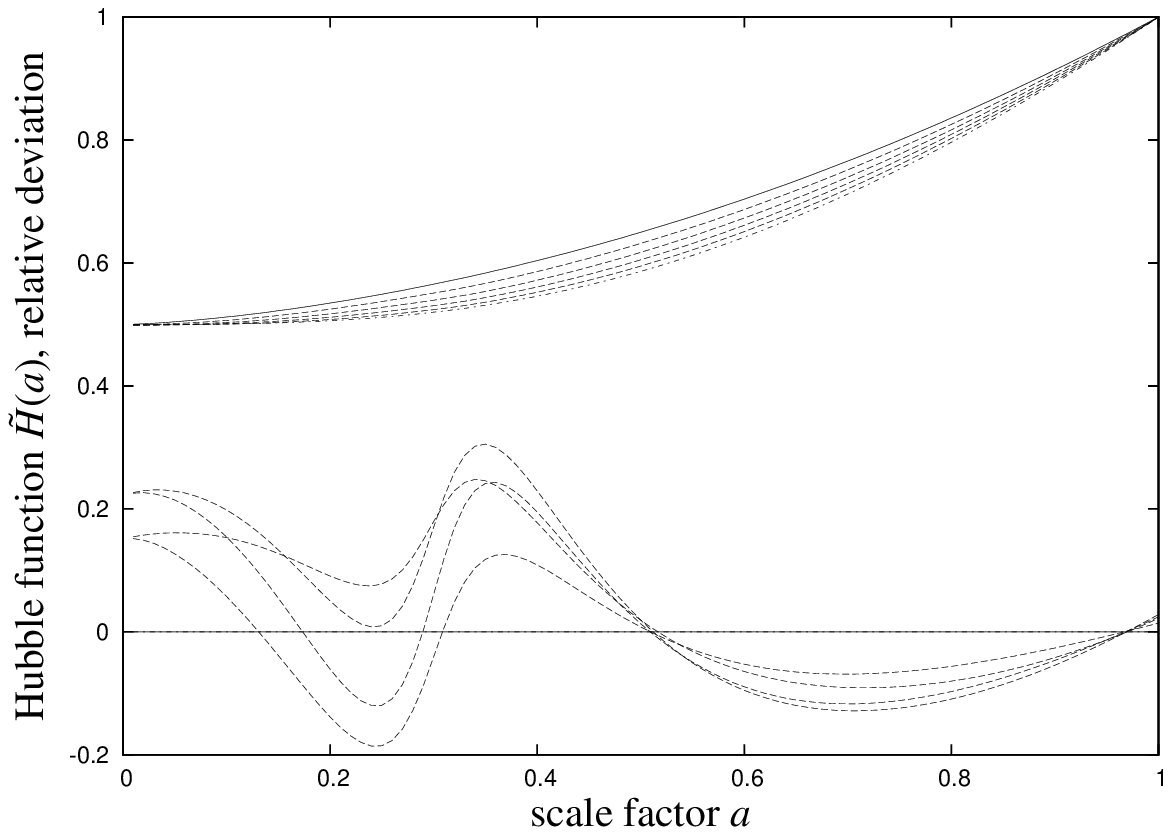}}
\caption{The scaled Hubble function $\tilde{H}(a)$ from the numerical approximation, as a function scale factor $a$, for values of $\alpha=0$ (corresponding to $\Lambda$CDM, dash-dotted), $\alpha=1$ (DGP-gravity, solid) and intermediate values in steps of $\Delta\alpha=0.2$ (dashed). At the bottom, the relative deviation {\em in percent} between the direct numerical solution for $H(a)$ and the approximation is plotted.}
\label{fig_hubble_approx}
\end{figure}

As discussed in Sect.~\ref{sect_homogeneous}, our family of modified Friedmann equations,
\begin{equation}
H^2(a) = (1-\Omega_m) H^\alpha(a) + \frac{\Omega_m}{a^3}
\label{appendix_eqn_friedmann}
\end{equation}
can only be solved numerically for $H(a)$. In order to have a much faster analytic approximation for the spherical collapse equations, we use two second order expansions, yielding quadratic equations, which are analytically solvable. At late times $a=1$, we expand $H^\alpha(a)$ at $H=1$ to second order,
\begin{equation}
H^\alpha\simeq 1 + \alpha (H - 1) + \frac{\alpha(\alpha-1)}{2} (H-1)^2,
\end{equation}
substitute into eqn.~(\ref{appendix_eqn_friedmann}), and solve the resulting quadratic equation for the approximation $H_+(a)$, which is valid at late times. At early times, where the Hubble function diverges, we expand $\tilde{H}_0(a)=a^\frac{3}{2}H(a)$, which assumes the value $\sqrt{\Omega_m}$ at $a=0$:
\begin{equation}
\tilde{H}^\alpha\simeq 
\Omega_m^{\frac{\alpha}{2}} + 
\alpha\Omega_m^{\frac{\alpha-1}{2}}(\tilde{H}-\sqrt{\Omega_m}) +
\frac{\alpha(\alpha-1)}{2}\Omega_m^{\frac{\alpha-2}{2}}(\tilde{H}-\sqrt{\Omega_m})^2.
\end{equation}
This expansion can be converted to a second order expansion for $H(a)$, substituted into eqn.~(\ref{appendix_eqn_friedmann}) and the resulting quadratic equation solved for the early time approximation $H_-(a)$. The relation
\begin{equation}
H(a) = \gamma_-(a)\:H_-(a) + \gamma_+(a)\:H_+(a)
\end{equation}
interpolates between the two branches by adjusting the weight $a$ accordingly, and avoids any sharp transition. $\gamma_\pm(a)$ is a nonlinear interpolation,
\begin{equation}
\gamma_\pm(a) = \frac{1}{2}\left[1\pm\frac{2}{\pi}\arctan\left(\frac{a-\mu}{\Delta}\right)\right]
\label{appendix_eqn_interpolation}
\end{equation}
with the transition scale factor $\mu=0.3$ and the transition width $\Delta=0.05$. Eqn.~(\ref{appendix_eqn_interpolation}) is normalised in the sense $\gamma_+(a)+\gamma_-(a)=1$. Figure~\ref{fig_hubble_approx} compares the approximation of the Hubble function to the direct numerical solution, and gives the relative deviation between the two: The approximation reproduces the Hubble function to an accuracy better than 0.3\%, which is accurate enough for the purpose of spherical collapse computations. It should be noted, that the cases $\alpha=0$ ($\Lambda$CDM) and $\alpha=1$ (DGP-gravity) are reproduced exactly.

% --- appendix: CMB spectra --- %
\section{Additional constraints from CMB temperature and polarisation spectra}\label{sect_appendix_cmb}
The constraints on modified gravity from cluster number counts can be significantly improved by adding priors on $\Omega_m$, $\sigma_8$ and $n_s$ from CMB temperature and polarisation spectra. The Fisher matrix of the CMB temperature and polarisation power spectra is given by \citep{1997ApJ...488....1Z, 2003ApJ...588L...5B}:
\begin{equation}
F_{\mu\nu}^\mathrm{CMB} = \sum_{\ell=2}^{\ell_\mathrm{max}}\sum_{X,Y}
\frac{\partial C_X(\ell)}{\partial x_\mu} \mathrm{Cov}^{-1}\left[C_X(\ell) C_Y(\ell)\right]\frac{\partial C_Y(\ell)}{\partial x_\nu},
\end{equation}
where the indices $(X,Y)$ assume values of $T$ for the temperature spectrum $C_T(\ell)$, $E$ for the $E$-mode polarisation spectrum $C_E(\ell)$ and $C$ for the temperature-polarisation cross-spectrum $C_C(\ell)$.

The entries of the covariance matrix $\mathrm{Cov}\left[C_X(\ell) C_Y(\ell)\right]$ are given by:
\begin{eqnarray}
\mathrm{Cov}\left[C_T(\ell) C_T(\ell)\right] & = & \frac{2}{2\ell+1}\frac{1}{f_\mathrm{sky}}\tilde{C}_T^2(\ell), \\
\mathrm{Cov}\left[C_E(\ell) C_E(\ell)\right] & = & \frac{2}{2\ell+1}\frac{1}{f_\mathrm{sky}}\tilde{C}_E^2(\ell), \\
\mathrm{Cov}\left[C_C(\ell) C_C(\ell)\right] & = & \frac{2}{2\ell+1}\frac{1}{f_\mathrm{sky}}\left(\tilde{C}_C^2(\ell)+\tilde{C}_T(\ell)\tilde{C}_E(\ell)\right), \\
\mathrm{Cov}\left[C_T(\ell) C_E(\ell)\right] & = & \frac{2}{2\ell+1}\frac{1}{f_\mathrm{sky}}\tilde{C}_C^2(\ell), \\
\mathrm{Cov}\left[C_E(\ell) C_C(\ell)\right] & = & \frac{2}{2\ell+1}\frac{1}{f_\mathrm{sky}}\tilde{C}_E(\ell)\tilde{C}_C(\ell),~\mathrm{and} \\
\mathrm{Cov}\left[C_T(\ell) C_C(\ell)\right] & = & \frac{2}{2\ell+1}\frac{1}{f_\mathrm{sky}}\tilde{C}_T(\ell)\tilde{C}_C(\ell),
\end{eqnarray}
with the observed spectra
\begin{eqnarray}
\tilde{C}_T(\ell) & = & C_T(\ell) + w_T^{-1}B^{-2}(\ell), \\
\tilde{C}_E(\ell) & = & C_E(\ell) + w_P^{-1}B^{-2}(\ell),
\end{eqnarray}
and $\tilde{C}_T(\ell)=C_C(\ell)$. For PLANCK's noise levels the values $w_T^{-1}=(0.02\umu\mathrm{K})^2$ and $w_P^{-1}=(0.03\umu\mathrm{K})^2$ have been used, and the beam was assumed to be Gaussian, $B^{-2}(\ell)=\exp(\Delta\theta^2\:\ell(\ell+1))$, with a FWHM-width of $\Delta\theta = 7\farcm1$, corresponding to the $\nu=143$~GHz channel closest to the CMB-maximum. The observation uses a fraction of $f_\mathrm{sky}=0.8$ of the entire sky, corresponding to a galactic cut excluding $\left|b\right|<11.5^\circ$.  CMB spectra and their derivatives were generated with CAMB\footnote{\tt http://camb.info/} \citep{2000ApJ...538..473L}.

\bsp

\label{lastpage}

\end{document}